\documentclass[12pt]{JHEP3}

 \preprint{  }

\usepackage{epsfig,multicol}
\usepackage{amsmath}
\usepackage{graphics}
\usepackage{graphicx}
\usepackage{dcolumn}
\usepackage{amssymb}

\title{Analytical study on holographic superconductors for Born-Infeld
electrodynamics in Gauss-Bonnet gravity with backreactions}
\author{
Weiping Yao, Jiliang  Jing\footnote{Corresponding author, Email:
jljing@hunnu.edu.cn}$^{1}$
\\ Institute of Physics and
Department of Physics, and Key Laboratory of Low Dimensional Quantum Structures and
Quantum Control of Ministry of Education, Hunan Normal University,
Changsha, Hunan 410081, P. R. China}

\abstract{
We analytically study the  holographic superconductors for Born-Infeld
electrodynamics in Gauss-Bonnet gravity with backreactions. We note
that the analytic method is still powerful for this complex system
and the results obtained by the analytical and numerical computations
are consistent. We find that the critical temperature decreases with
the increase of the backreactions, Gauss-Bonnet, and Born-Infeld
parameters, which means that increase of the strength of these
effects will make the scalar hair harder to form.
Furthermore, the Gauss-Bonnet factor modifies the
critical temperature more significantly than the backreaction factor.
The effect of the Born-Infeld factor on the critical temperature is weaker
than the backreaction factor. We also show
that the critical exponent is not affected by the backreactions,
Gauss-Bonnet gravity, and Born-Infeld electrodynamics.}


\begin{document}

\section{Introduction}

According to the anti-de Sitter/conformal field theories (AdS/CFT) correspondence
\cite{J.Maldacena1998, EWritten1998, S.S.Gubser1998}, the holographic model of superconductors is established
by a gravitational theory with a Maxwell field coupling to a charge scalar field
\cite{G.T.Horowitz2010, S.A.Hartnoll2009, C.P.Herzog2006}. Gubser
\cite{S.S.Gubser2005, S.S.Gubser2008} first suggested
that near the horizon of a charged black hole there is in operation a
geometrical mechanism parameterized by a charged scalar field of breaking
a local $U(1)$ gauge symmetry. Then, Hartnoll {\it{et al.}}
\cite{S.A.Hartnoll2008, S.A.Hartnoll2008j} found that the spontaneous $U(1)$
symmetry breaking by bulk black holes can be used to construct
gravitational dual of the transition from normal state to superconducting
state in the boundary theory, which exhibits the behavior of the superconductor.
Since this principle provides a new insight into the investigation of strongly
interacting condensed matter systems where the perturbational methods are no
longer available, there are a lot of works studying the AdS/CFT duality to
condensed matter physics and in particular to superconductivity
(for reviews, see refs. \cite{Ge-Wang-Wu,gregorysoda,Nakano-Wen,Amado,
Koutsoumbas,Umeh,Sonner,Jing-Chen,Franco,Herzog-2010,HorowitzPRD78,
 Konoplya,Siopsis,maeda,CaiNie,Pan-Wang,panwang,jingpanl} and references therein).

Recently, motivated by the application of the Mermin-Wagner theorem
to the holographic superconductors, it is of great interest to
study the effect of a particularly gravitational correction.
Among the gravity theories with higher curvature corrections,
the Gauss-Bonnet gravity attracted much attention because of
its several special features. For example, we can find the analytic expression of static,
spherically symmetric black hole solution in Gauss-Bonnet
gravity \cite{D.G.Boulware1985, J.T.Wheeler1986, R.G.Cai2002}.
To examine the effects of higher curvature corrections on the holographic superconductors,
Gregory {\it{et al.}}\cite{gregorysoda} analytically studied
the $(3+1)$-dimensional holographic superconductors in the Einstein-Gauss-Bonnet
gravity by matching method and found that the higher curvature corrections make
the scalar hair harder to form. However, a lot of studies on the holographic
superconductors focus on the probe approximation where the backreactions of
matter fields on the spacetime metric are neglected. When thinking about the
backreactions in the Einstein-Gauss-Bonnet gravity, Kanno \cite{Kanno:2011}
found that not only the higher curvature correction but also the backreactions
can make the condensation harder. Inspired by the curvature correction to gravity, it is of interest to study the
effects of higher order corrected Maxwell field on the scalar condensation, in
order to understand the influences of the 1/N corrections on the holographic
models, and compare them with the effects brought by the corrections to
gravity\cite{panjingwang1}. It is well known that
the Born-Infeld electrodynamics is a possible nonlinear electrodynamics
which carries more information than the Maxwell field and it has been a
subject of research for many years
 \cite{M.Born1934, G.W.Gibbons1995, B.Hoffmann1935, W.Heisenberg1936,  Oliveira1994, O.Miskovic2011}.
  Heisenberg and Euler \cite{W.Heisenberg1936}~noted
that quantum electrodynamics predicts that the electromagnetic field behaves non-linearly through
the presence of virtual charged particles. The Born-Infeld electrodynamics
proposed by Born and Infeld had the aim of obtaining a finite value for the self-energy of a point-like charge \cite{M.Born1934}.
The Lagrangian density for Born-Infeld theory is $L_{BI}=\frac{1}{b^2}\left(1-\sqrt{1+\frac{b^2F^{ab}F_{ab}}{2}}\right)$
with $F^2=F_{\mu\nu}F^{\mu\nu}$ and this Lagrangian will reduce to the Maxwell case as the coupling parameter $b$
approaches zero. The static spherically symmetric black holes for the Born-Infeld electrodynamics coupled
to Einstein gravity was studied in Refs. \cite{B.Hoffmann1935, Oliveira1994}.
Holographic superconductor models with Born-Infeld electrodynamics have
been employed in the study of its influence on the scalar condensation \cite{R.G.Cai2010,J.L.Jing2011,S.Ganopadhyay2011}.
Jing and Chen found that the Born-Infeld coupling parameter will make it harder for the scalar condensation to form \cite{J.L.Jing2010}.
Liu {\it {et al}} \cite{yunqi:2012} numerically studied the holographic
superconductor away from the probe limit by considering the corrections
both in the gravity and in the gauge matter fields. In order to gain more
insights in the effect of the curvature corrections, the nonlinear correction
in the electromagnetic fields and  the backreaction of matter fields on the
spacetime metric, in this paper we would like to analytically study the
holographic superconductor for the Born-Infeld electrodynamics in the
Gauss-Bonnet gravity with backreactions and discuss the effects of the
corrections and the backreactions on the condensation.

The structure of this paper is as follows. In Sec. II,
we will introduce the holographic superconductor models
and derive the equations of motions.  In Sec. III, we will
give an analytical investigation of the critical temperature
by using the matching method. In Sec. IV, we analytically
study the critical exponent. In Sec. V,  we will conclude
our main results of this paper.

\section{basic set up for holographic superconductors}

The action for a Born-Infeld electromagnetic field coupling
with a charged scalar field in the five-dimensional
Einstein-Gauss-Bonnet spacetime reads
\begin{eqnarray}\label{action}
S&=&\frac{1}{2\kappa^2}\int d^5
x\sqrt{-g}\left[R+\frac{12}{l^2}+\frac{\alpha}{2}
(R^{abcd}R_{abcd}-4R^{ab}R_{ab}+R^2)\right]\nonumber\\
&+&\int d^5 x\sqrt{-g}\left[\frac{1}{b^2}\left(1-\sqrt{1+\frac{b^2
F^{ab}F_{ab}}{2}}\right)+(-|\nabla\Psi-i q A\Psi
|^2-m^2|\Psi|^2)\right]\nonumber,\\
\end{eqnarray}
where $\kappa$ is the five dimensional
gravitational constant with $\kappa^2=8 \pi G_5$ and
$G_5$ is the five-dimensional Newton constant, $g$
is the determinant of the metric. $R$, $R_{\mu\nu\lambda\rho}$
and $R_{\mu\nu}$ are the Ricci scalar, Riemann curvature tensor,
and the Ricci tensor respectively. $l$ is
the AdS radius, $q$ and $m$ are respectively the
charge and the mass of the scalar field. $\alpha$, $b$ are
the Gauss-Bonnet coupling parameter and the Born-Infeld
coupling parameter respectively. Considering the causality constraint of the boundary CFT
\cite{Mauro Brigante2007, M.Brigante2008, A.Buchel2009, D.M.Hofman2010, J.de Boer2010,
X.O.Camanho2010, A.Buchel2010}, the range of parameter $\alpha$ must be restricted as
$\frac{-7l^2}{36}\leq\alpha\leq\frac{9l^2}{100}$.

In order to seek for a charged plane-symmetric hairy black
hole solution with backreactions, we take a metric ansatz in the form
\begin{eqnarray}\label{metricf}
ds^2=-f(r)e^{-\chi(r)}dt^2+\frac{dr^2}{f(r)}+\frac{r^2}{l_e^2}(dx^2+dy^2+dz^2)~.
\end{eqnarray}
The effective asymptotic AdS scale can be defined as
\begin{eqnarray}
l_e^2=\frac{l^2}{2}\left[1+\sqrt{1-\frac{4\alpha}{l^2}}\right].
\end{eqnarray}
Due to the spherical symmetry, the electromagnetic field and the scalar field can be chosen as
\begin{eqnarray}
&&A_\mu=(\phi(r), 0, 0, 0, 0), \nonumber \\ &&\Psi=\Psi(r),
\end{eqnarray}
without loss of generality, $\Psi(r)$ can be taken as a real function.
The Hawking temperature of the black hole is determined by
\begin{eqnarray}\label{temperature}
T_H=\left.\frac{f'(r)e^{-\chi(r)/2}}{4
\pi}\right|_{r=r_h}~.
\end{eqnarray}

Considering the ansatz of the metric (\ref{metricf}) and using
the action (\ref{action}), the equations of the motions are described by
\begin{eqnarray}\label{equationsofmotion}
&&\psi ''(r)+\psi '(r)\left[\frac{3}{r}+\frac{f'(r)}{f(r)}-\frac{\chi '(r)}{2}\right]+\psi
(r) \left[\frac{q^2 \phi (r)^2 e^{\chi
(r)}}{f(r)^2}-\frac{m^2}{f(r)}\right]=0, \\
&&\phi ''(r)+\left(\frac{\chi'(r)}{2}+\frac{3}{r}\right)\phi
'(r)-\frac{3b^2 e^{\chi (r)}}{r}\phi'(r)^3-\frac{2 q^2 \phi (r)\psi (r)^2}{f(r)}
(1-b^2 e^{\chi (r)}\phi '(r)^2 )^{\frac{3}{2}}=0,\nonumber  \\ \\
&&(1-\frac{2 \alpha f(r)}{r^2})\chi '(r)+\frac{4}{3} \kappa ^2 r
\left[\frac{q^2 \phi (r)^2 \psi (r)^2 e^{\chi(r)}}{f(r)^2}+\psi '(r)^2\right]=0, \\
&&(1-\frac{2 \alpha
f(r)}{r^2})f'(r)+\frac{2 f(r)}{r}-\frac{4r}{l^2}+\frac{2\kappa
^2 r }{3} \left[\left(\frac{q^2 \phi (r)^2 \psi (r)^2 e^{\chi
(r)}}{f(r)}+f(r) \psi '(r)^2\right.\right.\nonumber\\&&\left.\left.+m^2 \psi
(r)^2\frac{e^{\chi (r)} q^2
\phi(r)^2 \psi(r)^2}{f(r)}\right)+\frac{1}{b^2}\left[\left(1-b^2
\phi'(r)^2\right)^{-1/2}-1\right]\right]=0,
\end{eqnarray}
where the prime denotes the derivative with respect to $r$.
By respectively using the scaling symmetry
\begin{equation}\label{l:symmetry}
l\rightarrow a l,~\alpha\rightarrow a^2
\alpha,~r\rightarrow ar,~~q\rightarrow q
/a,~m^2\rightarrow m^2/a^{2},~b^2\rightarrow a^2b^2 ,
\end{equation}
\begin{equation}\label{q:symmetry}
q \rightarrow aq,~\phi \rightarrow \phi/a,~\psi
\rightarrow \psi/ a,~\kappa^2 \rightarrow
\kappa^2 a^{2},~b^2\rightarrow a^2 b,
\end{equation}
we can set the AdS radius and the charge of
scalar field as unity.

In order to obtain the solutions in superconducting phase,
we need to count on the appropriate boundary conditions.
At the horizon $r_+$, the metric functions $\chi$ and $f$ satisfy
\begin{eqnarray}
\chi^\prime(r_+)&=&-\frac{4\kappa^2}{3}r_+\left(
\psi^\prime(r_+)^2
+\frac{q^2\phi^\prime(r_+)^2\psi(r_+)^2e^{\chi(r)}}{f^\prime(r_+)^2}
\right), \\
f^\prime(r_+)&=&\frac{4}{l^2}r_+-\frac{2\kappa^2r_+}{3}\bigg(m^2\psi
(r_+)^2\frac{e^{\chi (r_+)}q^2
\phi(r_+)^2\psi(r_+)^2}{f(r_+)}\nonumber \\
&+&\frac{1}{b^2}\left[\left(1-b^2
\phi'(r_+)^2\right)^{-1/2}-1\right]\bigg),
\end{eqnarray}
and  $\phi$ and $\psi$ are given by
\begin{eqnarray}
\phi(r_+)=0, \hspace{1cm}
\psi^\prime(r_+)=\frac{m^2}{f^\prime(r_+)}\psi(r_+).
\end{eqnarray}
As the spacetime is asymptotically AdS,
 the asymptotic behaviors of the solutions at the spatial infinity are
\begin{eqnarray}\label{b:infinity}
\chi\rightarrow0\, , \hspace{0.5cm}
f\sim\frac{r^{2}}{l^{2}}\, , \hspace{0.5cm}
\phi\sim\mu-\frac{\rho}{r^{2}}\, , \hspace{0.5cm}
\psi\sim\frac{\psi_{-}}{r^{\Delta_{-}}}+\frac{\psi_{+}}{r^{\Delta_{+}}}\, ,
\label{infinity}
\end{eqnarray}
where $\Delta_\pm=2\pm\sqrt{4+m^2l_e^2}$, $\mu$ and $\rho$ are the
chemical potential and charge density. According to the AdS/CFT
correspondence, the coefficients $\psi_-$ and $\psi_+$ correspond
to the vacuum expectation values
$\psi_{-}=<\mathcal{O}_{-}>$,  $\psi_{+}=<\mathcal{O}_{+}>$ of
an operator $\mathcal{O}$ dual to the scalar field. For the sake
of obtaining stability in the asymptotic AdS region, we can impose
boundary conditions that either $\psi_-$ or $\psi_+$ vanishes.
In the following calculation, we will focus on the condition $\psi_-=0$.

\section{critical temperature}

In this section, we will analytically study the critical temperature for
the Born-Infeld electrodynamics in the Gauss-Bonnet gravity with backreactions.

Let us set $z=r_+/r$, then the equations of the motions become
\begin{eqnarray}
&&\psi^{\prime\prime}+\left(\frac{f\prime}{f}
-\frac{\chi^\prime}{2}-\frac{1}{z}\right)\psi^\prime
+\frac{r_+^2}{z^4}\left(
\frac{q^2\phi^2e^{\chi}}{f^2}-\frac{m^2}{f}\right)\psi=0,
\label{z:psi}\\
&&\phi^{\prime\prime}+(\frac{\chi^\prime}{2}-\frac{1}{z})\phi^\prime
+\frac{3b^2~e^{\chi}}{r_+^2}z^3\phi'^3-\frac{2 q^2 \phi\psi^2r_+^2}{fz^4}
(1-\frac{b^2~e^{\chi}\phi '^2z^4}{r_+^2})^{\frac{3}{2}}=0,
\label{z:phi}\\
&&\left(1-2\alpha\frac{z^2}{r_+^2}f\right)\chi^\prime
-\frac{4\kappa^2}{3}\frac{r_+^2}{z^3}\left(
\frac{q^2\phi^2\psi^2e^{\chi}}{f^2}+\frac{z^4}{r_+^2}\psi^{\prime2}
\right)=0,
\label{z:chi}\\
&&\left(1-2\alpha\frac{z^2}{r_+^2}f\right)f^\prime
-\frac{2}{z}f+\frac{4r_+^2}{l^2z^3}-\frac{2\kappa^2}{3}\frac{r_+^2}{z^3}
\left[\left(\frac{q^2\phi^2\psi^2e^{\chi}}{f}+\frac{f\psi'^2z^4}{r_+^2}
+m^2\psi^2\right.\right.\nonumber\\&&\left.\left.\frac{e^{\chi}q^2
\phi^2\psi^2}{f}\right)-
\frac{1}{b^2}(1-(1-\frac{b^2z^4\phi'^2}{r_+^2})^{-1/2})\right]=0,
\label{z:f}
\end{eqnarray}
where the prime now denotes a derivative with respect to $z$.
The region $r_+<r<\infty$ now corresponds to $1>z>0$. Since
the value of the scalar operator $<\mathcal{O}_{+}>$ is small
near the critical point, we can use it as an expansion parameter
\begin{eqnarray}
\epsilon\equiv\langle{\cal O}_{\Delta_+}\rangle\, .
\end{eqnarray}
Because $\psi$ is small near the critical point,
we can expand the scalar field $\psi(z)$ and the gauge field $\phi(z)$ as follows
\cite{Herzog-2010,Kanno:2011, Ge2011}:
\begin{eqnarray}
\psi&=&\epsilon\psi_1+\epsilon^3\psi_3+\epsilon^5\psi_5+\cdots,\label{e:psi} \\
\phi&=&\phi_0+\epsilon^2\phi_2+\epsilon^4\phi_4+\cdots,
\end{eqnarray}
and the metric function $f(z)$ and $\chi(z)$ can be expanded around the Gauss-Bonnet-AdS spacetime
\begin{eqnarray}
f&=&f_0+\epsilon^2f_2+\epsilon^4f_4+\cdots ,  \\
\chi&=&\epsilon^2\chi_2+\epsilon^4\chi_4\cdots. \label{e:chi}
\end{eqnarray}
For the chemical potential $\mu$,   we will allow it to be corrected order by order \cite{Herzog-2010}
\begin{eqnarray}
\mu=\mu_0 + \epsilon^2\delta\mu_2\, ,
\end{eqnarray}
where $\delta\mu_2$ is also positive.
Obviously, the parameter $\epsilon$ becomes zero when $\mu=\mu_0$,
which means the critical value of $\mu$ is $\mu_c=\mu_0$.
Now we begin to solve the equations order by order.

For zeroth order,  the Eqs. (\ref{z:phi}) and (\ref{z:f}) become
\begin{eqnarray}
&&\phi_0^{\prime\prime}-\frac{1}{z}\phi_0^\prime
+\frac{3b^2}{r_+^2}z^3\phi_0'^3=0, \label{z:phi0} \\
&&(1-\frac{2\alpha f_0}{r_+^2}z^2)f'_0-\frac{2f_0}{z}
+\frac{4r_+^2}{l^2z^3}+\frac{2\kappa^2}{3}\frac{r_+^2}{z^3}
\frac{1}{b^2}(1-(1-\frac{b^2z^4\phi_0'^2}{r_+^2})^{-1/2})=0. \label{z:f0}
\end{eqnarray}
Introducing a new function~$\lambda=\phi_0'^2$~, the Eq.~(\ref{z:phi0}) can be rewritten as
\begin{equation}\label{lamda}
\lambda'-\frac{2}{z}\lambda+\frac{3b^2}{r_+^2}z^3\lambda^2=0.
\end{equation}
Consequently, the solution of the Eq.~(\ref{lamda}) is
\begin{equation}\label{phi:lambda}
\lambda=\frac{4r_+^2z^2\mu_0^2}{r_+^2+2b^2z^6\mu_0^2}.
\end{equation}
Noting the coupling parameter~$0<b^2<<1$ and  using the boundary
condition $f(r_+)=0$, the solution of Eq. (\ref{z:f0}) is given by
\begin{eqnarray}\label{f0}
f_0(z)=\frac{r_+^2}{2\alpha z^2}\left[
1-\sqrt{1-\frac{4\alpha}{l^2}(1-z^4)
+2\kappa^2\frac{4\alpha\mu_0^2}{3r_+^2}z^4(1-z^2)+2\kappa^2b^2
\frac{\alpha\mu_0^4}{3r_+^4}z^4(1-z^8)}~\right].\nonumber \\
\end{eqnarray}
For the first order, the Eq.~(\ref{z:psi}) becomes
\begin{equation}\label{Z:psi1}
\psi_1^{\prime\prime}+\left(\frac{f_0^\prime}{f_0}
-\frac{1}{z}\right)\psi_1^\prime+\frac{r_+^2}{z^4}\left(
\frac{q^2\phi_0^2}{f_0^2}-\frac{m^2}{f_0}\right)\psi_1=0.
\end{equation}
The behavior of $\psi$ near the horizon $z\rightarrow1$ is
\begin{eqnarray}
\psi_1^\prime(1)=\frac{r_+^2m^2}{f_0^\prime(1)}\psi_1(1),
\label{1:regularity}
\end{eqnarray}
and near the boundary $z\rightarrow0$, we have
\begin{eqnarray}
\psi_1\sim D_+z^{\Delta_+}\, .
\label{1:boundary}
\end{eqnarray}
Since it is very difficult to deal with both of the backreaction on
the Gauss-Bonnet spacetime and the Born-Infeld term, we will use
the matching method~\cite{gregorysoda} to derive an analytic
expression for the critical temperature.

Near the horizon, the scalar field $\psi_1$ can be expanded as
\begin{eqnarray}
\psi_1(z)=\psi_1(1)-\psi_1^\prime(1)(1-z)
+\frac{1}{2}\psi_1^{\prime\prime}(1)(1-z)^2
+\cdots,
\end{eqnarray}
without loss of generality, we can take $\psi_1(1)>0$ to insure
$\psi_1(z)$ positive. Then, we find that the second derivative
of the function $\psi_1(z)$ at the horizon can be expressed by
\begin{equation}\label{{S;secandpsi}}
\psi_1^{\prime\prime}(1)=\frac{1}{2}\left(
-3-\frac{f_0^{\prime\prime}(1)}{f_0^\prime(1)}+\frac{r_+^2m^2}{f_0^\prime(1)}
\right)\psi_1^\prime(1)
-\frac{q^2r_+^2\lambda(1)}{2f_0^\prime(1)^2}\psi_1(1).
\end{equation}
After eliminating $\psi_1^\prime(1)$ from above equation
by using Eq.~(\ref{1:regularity}),  the solution near the
horizon is
\begin{eqnarray}
\psi_1(z)&=&\psi_1(1)\left\{1-\frac{r_+^2m^2}{f_0^\prime(1)}(1-z)
+\left[-\frac{r_+^2m^2}{4f_0^\prime(1)}\left(
3+\frac{f_0^{\prime\prime}(1)}{f_0^\prime(1)}
-\frac{r_+^2m^2}{f_0^\prime(1)}
\right) \right. \right. \nonumber \\
&&\left. \left.-\frac{q^2r_+^2}{4}\frac{\lambda(1)}{f_0^\prime(1)^2}
\right](1-z)^2
+\cdots\right\},
\label{1:horizon}
\end{eqnarray}
where $\psi_1(1)$ is another unknown constant.

Now we try to connect the solutions Eqs. (\ref{1:boundary}) and
(\ref{1:horizon}) smoothly at $z_m$ and in order to determined the
constants $\psi_1(1)$ and $D_+$.
We require the following conditions
\begin{eqnarray}
z_m^{\Delta_+}D_+
&=&\psi_1(1)\left\{1-\frac{r_+^2m^2}{f_0^\prime(1)}(1-z_m)
-\left[
\frac{r_+^2m^2}{4f_0^\prime(1)}\left(
3+\frac{f_0^{\prime\prime}(1)}{f_0^\prime(1)}
-\frac{r_+^2m^2}{f_0^\prime(1)}
\right)\right.\right.
\nonumber\\ &&\left.\left. +\frac{q^2r_+^2}{4}\frac{\lambda(1)}{f_0^\prime(1)^2}
\right](1-z_m)^2\right\}, \label{m:psi}\,\nonumber  \\
\Delta_+ z_m^{\Delta_+-1} D_+
&=&\psi_1(1)\left\{\frac{r_+^2m^2}{f_0^\prime(1)}
-2\left[
-\frac{r_+^2m^2}{4f_0^\prime(1)}\left(
3+\frac{f_0^{\prime\prime}(1)}{f_0^\prime(1)}
-\frac{r_+^2m^2}{f_0^\prime(1)}
\right)\right.\right.
\nonumber\\ &&\left.\left.
-\frac{q^2r_+^2}{4}\frac{\lambda(1)}{f_0^\prime(1)^2}
\right](1-z_m)\right\}.
\label{m:dpsi}
\end{eqnarray}
From which we find the following relation between $\psi_1(1)$ and $D_+$ is
\begin{eqnarray}
D_+=\frac{2z_m^{1-\Delta_+}}{2z_m+(1-z_m)\Delta_+}\left(
1-\frac{1-z_m}{2}\frac{r_+^2m^2}{f_0^\prime(1)}
\right)\psi_1(1)\, .
\label{D}
\end{eqnarray}
To get a non-trivial solution ($\psi_1(1)\neq 0$), by plugging
the above constraint back into Eq.~(\ref{m:dpsi}), we find+6
\begin{eqnarray}
&&2E(z_m)f_0^\prime(1)^2
-\left(2E((z_m)F(z_m)+3F(z_m)+1\right)r_+^2m^2f_0^\prime(1)
\nonumber \\&&-F(z_m)r_+^2m^2f_0^{\prime\prime}(1)
+F(z_m)r_+^4m^4-F(z_m)r_+^2q^2\lambda(1)=0,
\label{nontrivial}
\end{eqnarray}
where
\begin{eqnarray}
E=\frac{\Delta_+}{2z_m+(1-z_m)\Delta_+}, \hspace{0.3cm}
F=\frac{1-z_m}{2}.
\end{eqnarray}
And from the Eq.~(\ref{f0}), we can achieve
\begin{eqnarray}\label{function}
&&f_0'(1)=\frac{4}{3}\bigg(-\frac{3r_+^2}{l^2}+\kappa^2\mu_0^2
(1+\frac{b^2\mu_0^2}{r_+^2})\bigg),\label{1function}\nonumber \\
&&f_0''(1)=\frac{4(9l^2r_+^8+3\kappa^2l^4r_+^4\mu_0^2(5r_+^2+11b^2\mu_0^2)
+8\alpha(\kappa^2l^2r_+^2\mu_0^2-3r_+^4
+b^2\kappa^2l^2\mu_0^4)^2)}{9l^4r_+^6}\label{function2}.\nonumber \\
\end{eqnarray}
Due to both of the parameters $\kappa^2$ and $b^2$ are small, the above
functions are expanded by $\kappa^2$ and $b^2$. Then, the function can be simplified as
\begin{equation}\label{sm:function}
 b^2A\mu_0^4+r_+^2B\mu_0^2-\frac{r_+^4}{l^4}C=0,
\end{equation}
with
\begin{eqnarray}\label{ABC}
&&A=16GE+(14m^2l^2G-16\alpha m^2G-8q^2z_m^6)F+2m^2l^2EFG+l^2m^2G \label{A},\nonumber \\
&&B=16GE+(8m^2l^2G-16\alpha m^2G+4q^2)F+2m^2l^2EFG+l^2m^2G \label{B},\nonumber  \\
&&C=32E+(m^4l^4-32\alpha m^2+8m^2l^2)F+2l^2EF+4m^2l^2\label{C}, \nonumber \\
&&G=\frac{4\kappa^2}{3l^2}.\label{G}
\end{eqnarray}

\subsection{Case for b=0}

For the $b=0$, the Born-Infeld field reduces to the Maxwell field,
and the solution of the Eq. (\ref{sm:function}) becomes
\begin{eqnarray}
\mu_0^2=\frac{r_+^2C}{l^4B},
\end{eqnarray}
thus, the critical temperature $T_c$ is given by
\begin{eqnarray}\label{temperature0}
T_c=\frac{\rho^{1/3}}{\pi l^{4/3}}(\frac{B}{C})^{1/6}\big( 1-\frac{\kappa^2C}{3 l^2B}\big).
\end{eqnarray}
Now, we have presented a complete picture of the critical temperature $T_c$ with the
Gauss-Bonnet holographic superconductors in Maxwell electrodynamics with backreactions.
\FIGURE{
\includegraphics[scale=0.43]{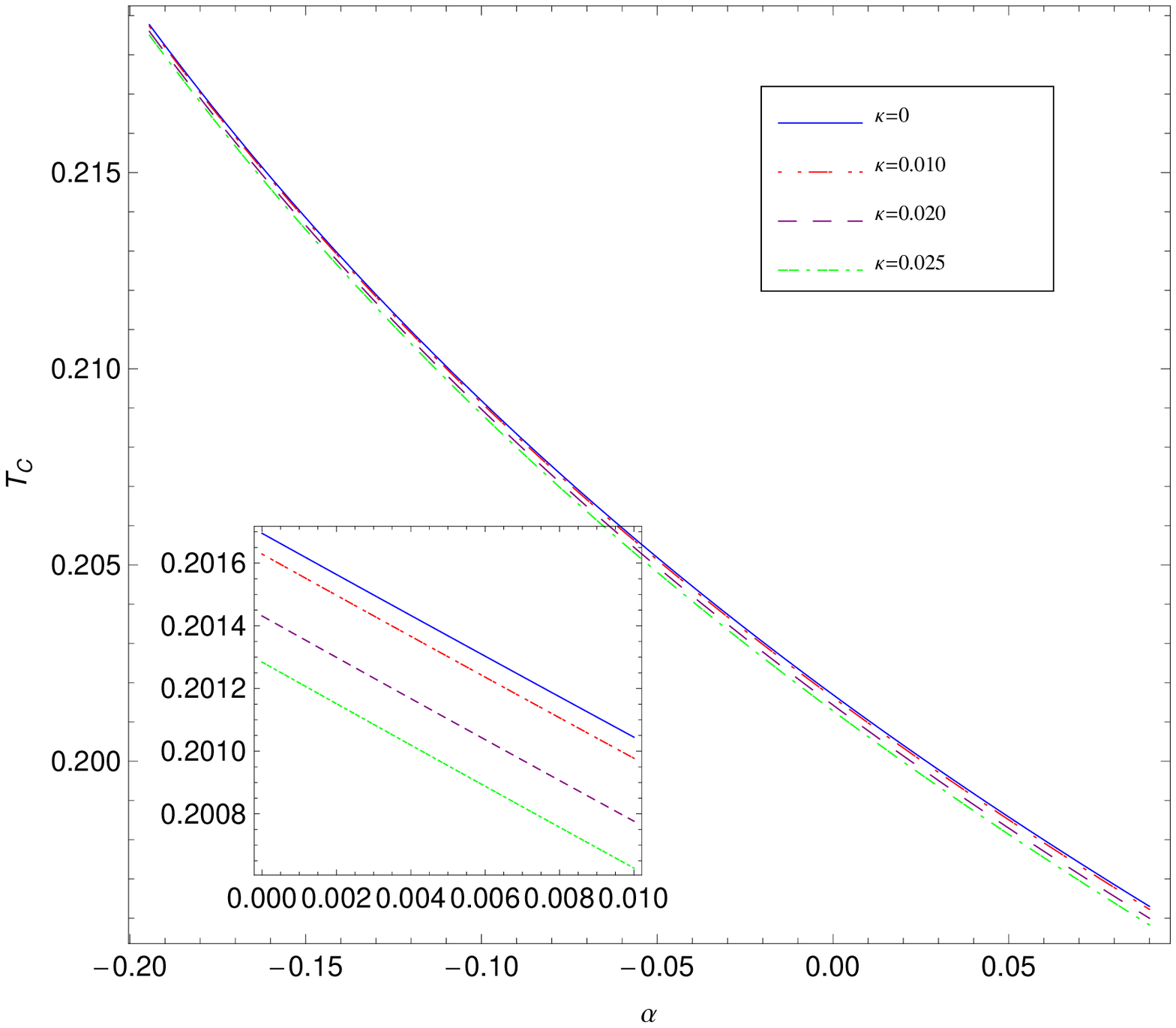}\hspace{0.2cm}%
\caption{The figure shows the critical temperature $T_c$ for the different
parameters~$\alpha$~and~$\kappa$.
 Here, we set $q=1$, $l=1$, $m^2l^2=-3$, $z_m =1/2$, $\rho=1$.}}
The critical temperature $T_c=0.201$ for the factors $\kappa=\alpha=0$ ,
$m^2l^2=-3$ and the matching point $z_m=1/2$, which agrees well with the
numerical result $T_c=0.198$ for $m^2l^2=-3$ in ref. \cite{Herzog-2010}. And
the analytical result for $m^2l^2=-4$,  $\kappa=\alpha=0$ by taking the matching point as $z_m=1/2$ is
also consistent with result in ref. \cite{Siopsis}.
In Fig. $1$, it is shown that, with the increase of the Gauss-Bonnet
factor $\alpha$ for fixed factor $\kappa$, the critical temperature decreases which
means that the Gauss-Bonnet factor $\alpha$ can make the scalar hair harder to form.
As the Gauss-Bonnet factor $\alpha$ is fixed, the critical temperature decreases
with the increase of the backreactions parameter $\kappa$. This means the backreactions factor can make the scalar condensation harder.

\subsection{Case for $b\neq 0$}

For $b\neq0$,  the solution of the Eq. (\ref{sm:function}) is
\begin{equation}\label{sm:solution}
\mu_0^2=\frac{r_+^2}{l^2}\frac{\sqrt{l^4B^2+4b^2AC}-Bl^2}{2b^2A},
\end{equation}
thus, the critical temperature is
\begin{equation}\label{c:Temperature}
T_c=\frac{\rho^{\frac{1}{3}}l^{-\frac{5}{3}}}{\pi}
\bigg(\frac{\sqrt{l^4B^2+4b^2AC}-Bl^2}{2b^2A}\bigg)^{-\frac{1}{6}}
\bigg(
1-\kappa^2\frac{\sqrt{l^4B^2+4b^2AC}-Bl^2}{2b^2A}\bigg).
\end{equation}
Obviously, the critical temperature $T_c$ depends on the parameters
$\alpha$, $b$ and $\kappa$. The analytical result is good agreement
with the numerical result for the Born-Infeld field in \cite{yunqi:2012}
for $\kappa=\alpha=0$ by taking $z_m=1/2$.

In Fig.$2$,  it is shown that, as the parameters $\kappa$ and $b$ are fixed,
the critical temperature decreases with the
increase of the Gauss-Bonnet parameter, which means
that the Gauss-Bonnet factor $\alpha$ makes the
condensation harder to form. As
the Born-Infeld parameter $b$ increases for fixed parameters
$\alpha$ and $\kappa$, the critical temperature decreases.
This means the scalar condensation can be more difficult to form for larger
Born-Infeld parameter. With the increase of
the backreaction parameter $\kappa$ for fixed parameters
$\alpha$ and $b$, the critical temperature decreases. This
means the scalar condensation becomes difficult as the
backreaction parameter $\kappa$ increases.
\FIGURE{
\includegraphics[scale=0.36]{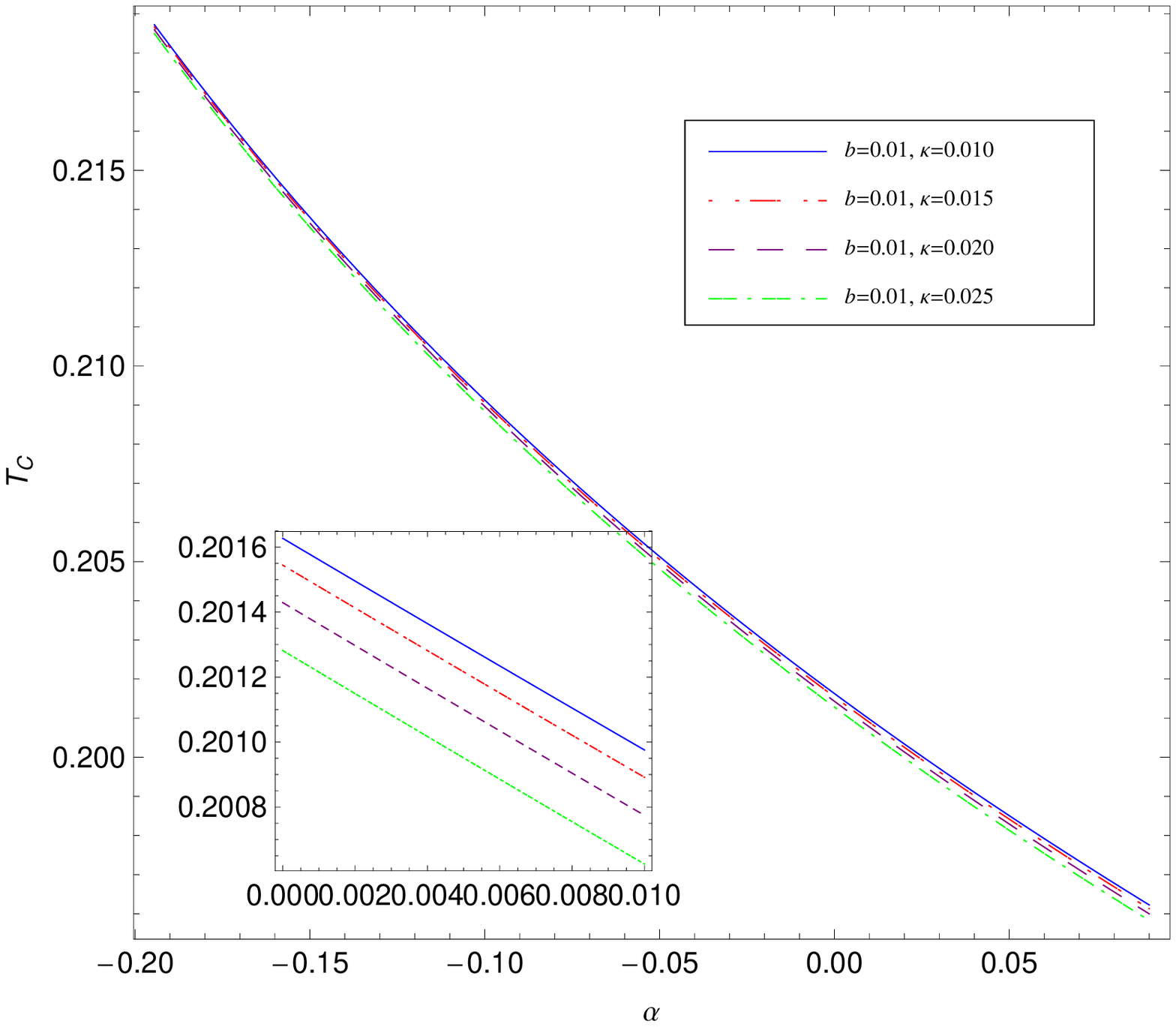}\hspace{0.2cm}%
\includegraphics[scale=0.50]{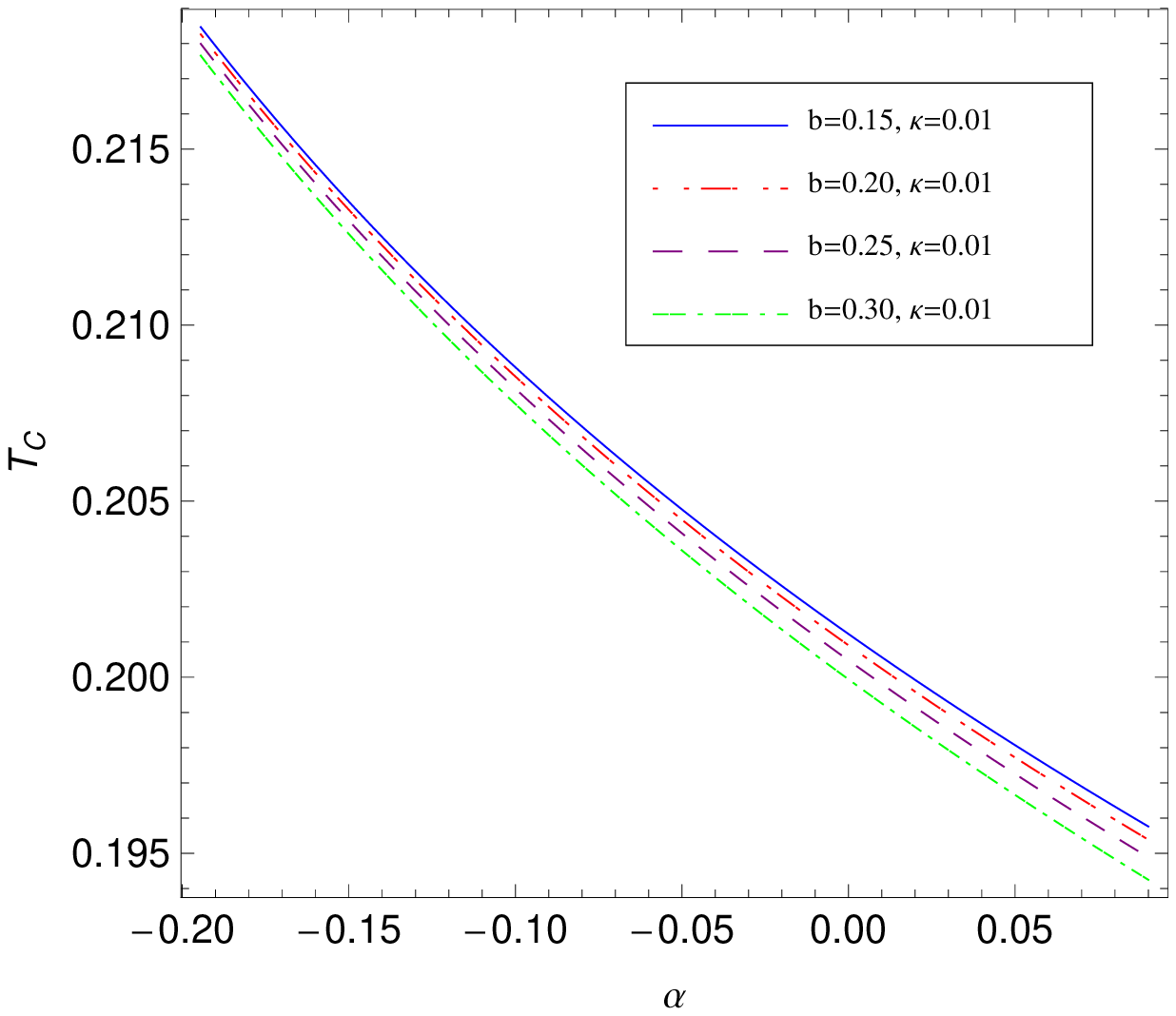}\hspace{0.2cm}%
\caption{The figures show the critical temperature $T_c$ for the different
parameters~$\alpha$,~$b$~and~$\kappa$.
Here,  we set  $q=1$, $l=1$, $m^2l^2=-3$, $z_m =1/2$, $\rho=1$.}}

In order to have a meaningful
notion for the critical temperature $T_c$ as the parameter $\alpha$ in the range of
$\frac{-7}{36}\leq\alpha\leq\frac{9}{100}$ with $l=1$, from Eqs. (3.31) and (3.32), we find that
the Born-Infeld parameter $b$ must satisfy
\begin{equation}\label{b}
0\leq b\leq\sqrt{\frac{1+26.12\kappa^2+170.61\kappa^4}{2.27-513.46\kappa^2}},
\end{equation}
and the backreaction parameter $\kappa$ is restricted as
\begin{equation}\label{k}
0\leq\kappa <0.067.
\end{equation}
On the other hand, to keep the
differences between the analytical and numerical values under $5\%$,
the range for backreaction parameter $\kappa$ should be $0\leq\kappa\leq 0.026$.

\begin{figure}
\includegraphics[scale=0.37]{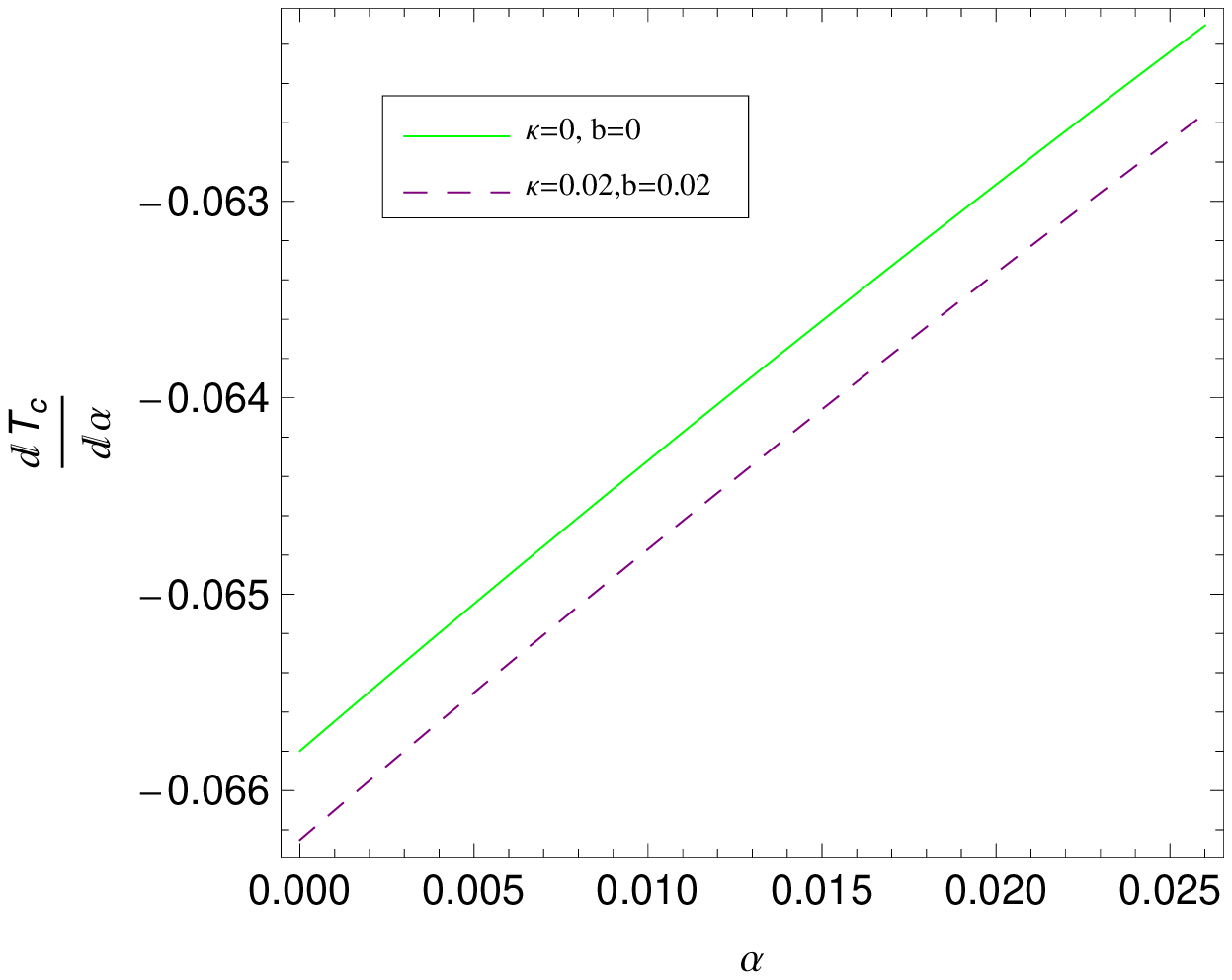}\hspace{0.2cm}%
\includegraphics[scale=0.37]{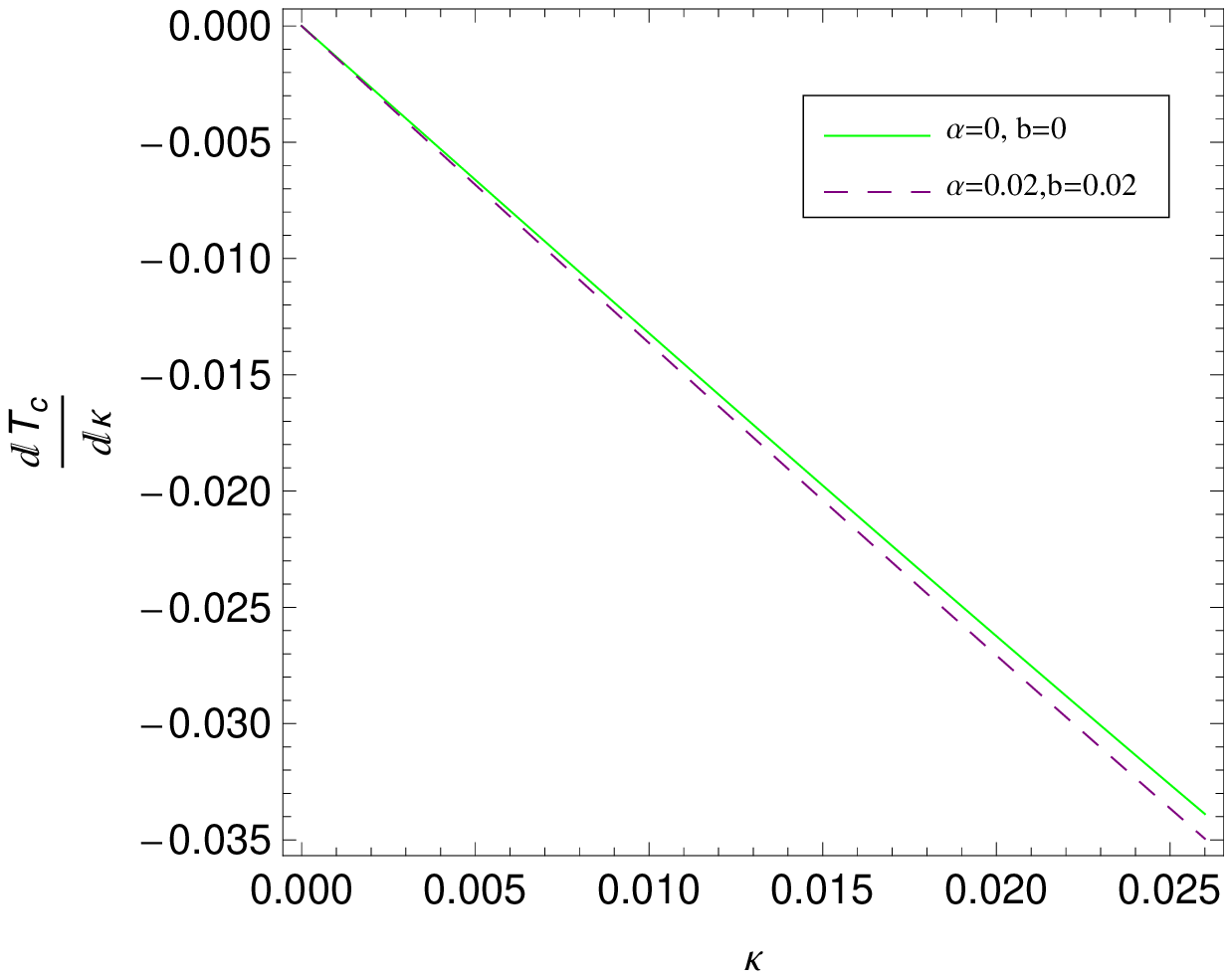}\hspace{0.2cm}%
\includegraphics[scale=0.37]{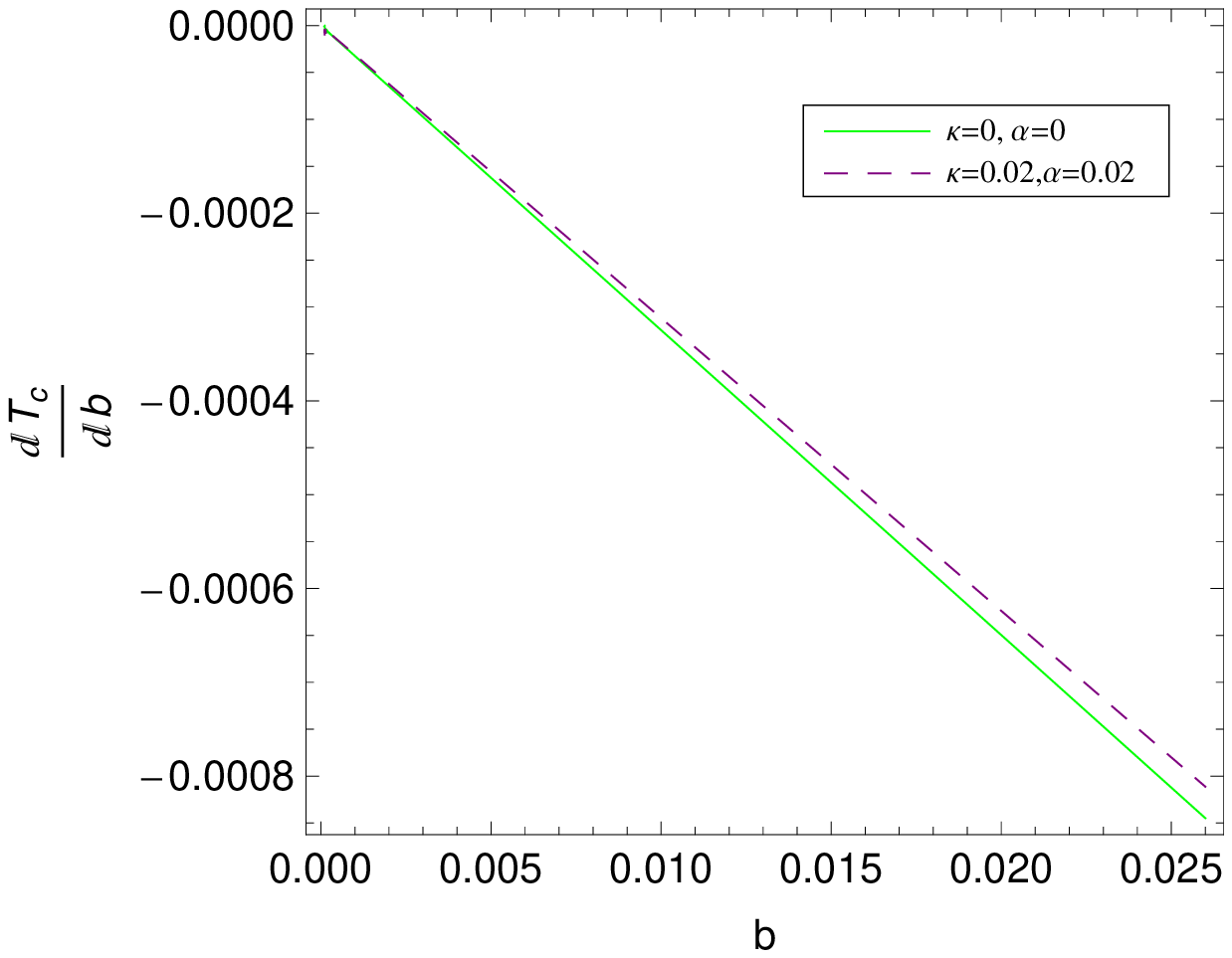}\hspace{0.2cm}%
\caption{The figures show the derivative of temperature critical $T_c$ with respect to
parameters~$\alpha$,~$\kappa$~and~$b$~respectively.
Here,  we set  $q=1$, $l=1$, $m^2l^2=-3$, $z_m =1/2$, $\rho=1$.}\label{dd}
\end{figure}

From Fig. \ref{dd}, we find that the value of
$\frac{\partial T_c }{\partial\alpha}$ increases with the increase of parameter $\alpha$, but its value is negative. That is to say,
the critical temperature $T_c$ decreases more slowly as the parameter
$\alpha$ increases. With the increase of parameters
$\kappa$ and $b$, the values of $\frac{\partial T_c }{\partial\kappa}$ and
$\frac{\partial T_c }{\partial b}$ decrease and their values are negative. This means that the critical temperature $T_c$
decreases faster as the parameters $\kappa$ and $b$ increase. We also find that the maximum value of
$\frac{\partial T_c }{\partial\alpha}\mid_{\kappa,b}$ is
less than the minimum value of $\frac{\partial T_c }{\partial\kappa}\mid_{\alpha,b}$,
which means the Gauss-Bonnet parameter $\alpha$ modifies the
critical temperature more significantly than the backreaction parameter $\kappa$. Similarly, we can find that the effect
of the backreaction parameter $\kappa$ on the critical temperature is bigger than the Born-Infeld parameter $b$.

In table I, we compare the analytical results with the numerical ones by
choosing different values for the backreaction parameter $\kappa$,
Born-Infeld parameter $b$ and Gauss-Bonnet parameter $\alpha$.
From the table, we find that the analytical results are consistent
with the numerical results. We also arrive at the conclusion that
the critical temperature decreases with the increase of the Gauss-Bonnet,
the backreactions and Born-Infeld parameters although there are
interactions between the parameters, which means that increase of the strength of
these effects will make the scalar hair harder to form.
\begin{center}
\begin{table}[ht]
\caption{
The analytical and numerical results of the critical temperature $T_c$ with
the chosen values of the backreaction parameter $\kappa$, Born-Infeld parameter
$b$ and Gauss-Bonnet parameter $\alpha$. Here,
we set $q=1$, $l=1$, $m^2l^2=-3$, $z_m =1/2$, $\rho=1$.}\label{table}
\begin{tabular}{c|c|c|c|c|c|c|c|c|c}\hline
\multicolumn{10}{c}{ Analytical results } \\
\hline
&\multicolumn{3}{c|}{$\alpha=-0.1$}
&\multicolumn{3}{c|} {$ \alpha=0$}
&\multicolumn{3}{c}{$\alpha=0.09$} \\
\hline
$\kappa$ $\Big\backslash b$ &0 &0.1 & 0.3 &0.1 & 0.2&0.5&0.1&0.4&0.6\\
\hline
$0.0$    &0.2092&0.2091 &0.2079 & 0.2015 & 0.2010&0.1967&0.1961&0.1928&0.1856\\
\hline
$0.01$  &0.2091&0.2090&0.2078&0.2014&0.2009&0.1962&0.1960&0.1924&0.1845\\
\hline
$0.02$  &0.2090&0.2088&0.2073&0.2012&0.2005&0.1948&0.1957&0.1913&0.1818\\
\hline
\multicolumn{10}{c}{ Numerical results } \\
\hline
&\multicolumn{3}{c|}{$\alpha=-0.1$}
&\multicolumn{3}{c|} {$ \alpha=0$}
&\multicolumn{3}{c}{$\alpha=0.09$} \\
\hline
$\kappa$ $\Big\backslash b$ &0 &0.1 & 0.3 &0.1 & 0.2&0.5&0.1&0.4&0.6\\
\hline
$0.0$    &~0.2090~&~0.2090~&~0.2090 ~&~ 0.1980 ~& 0.1980~&~0.1980~&~0.1874~&~0.1874~&~0.1874\\
\hline
$0.01$  &0.2089&0.2088&0.2087&0.1978&0.1977&0.1974&0.1872&0.1867&0.1860\\
\hline
$0.02$  &0.2087&0.2086&0.2082&0.1975&0.1972&0.1955&0.1867&0.1846&0.1818\\
\hline
\end{tabular}
\end{table}
\end{center}

\section{critical exponent}

The critical exponent of the Gauss-Bonnet holographic superconductors
for Born-Infeld electrodynamics with backreactions will be calculated
in the following. To investigate the critical exponent, we expand $\phi_0$
in a Taylor series at the horizon as
\begin{equation}\label{c:phi0}
\phi_0(z)=\phi_0(1)-\phi_0^\prime(1)(1-z)
+\frac{1}{2}\phi_0^{\prime\prime}(1)(1-z)^2+...
\end{equation}
Then, the approximate solution of the $\phi_0$ at the horizon is given by
\begin{eqnarray}\label{cs:phi01}
\phi_0(z)&=&\phi_0(1)-\phi_0^\prime(1)(1-z)+\frac{1}{2}\bigg[\phi_0^\prime(1)
-\frac{3b^2}{r_+^2}\phi_0^\prime(1)^3
\nonumber\\
&&+\frac{2 q^2 \phi'_0(1) \psi_1(1)^2r_+^2}{f'_0(1)}
(1-\frac{3b^2}{2 r_+^2}\phi_0^\prime(1)^2)\bigg](1-z)^2.
\end{eqnarray}
By matching it with the solution near the horizon smoothly at $z=z_m$, we have
\begin{eqnarray}
\mu_0-\frac{\rho z_m^2}{r_+^2}&=&\beta(1-z_m)-\frac{\beta}{2}\Big[1
-\frac{3b^2}{r_+^2}\beta^2+\frac{2 q^2\gamma^2r_+^2}{f'_0(1)}
(1-\frac{3b^2}{2 r_+^2}\beta^2)\Big](1-z)^2,\nonumber \\
\frac{2\rho z_m }{r_+^2}&=&\beta-\beta(1-z_m)\Big[1
-\frac{3b^2}{r_+^2}\beta^2+\frac{2 q^2\gamma^2r_+^2}{f'_0(1)}
(1-\frac{3b^2}{2 r_+^2}\beta^2)\Big]\label{c:phi0zm},
\end{eqnarray}
where $\beta=-\phi_0^\prime(1)$ and $\gamma=\psi_1(1)$.
From the Eq.~(\ref{c:phi0zm}) we obtain
\begin{equation}\label{c:psi1}
\gamma^2=\frac{1}{2q^2(1-z_m)}\frac{f'_0(1)}{r_+^2}\bigg[(1-\frac{2\rho }{\beta r_+^2})z_m
-\frac{3b^2\beta^2}{2r_+^2}(\frac{2\rho z_m}{\beta r_+^2}+z_m-2)\bigg].
\end{equation}
By using Eqs. (\ref{temperature}) and (\ref{1function}), the above equation becomes
\begin{eqnarray}\label{ct2:psi1}
\psi_1(1)^2=\frac{1}{q^2(1-z_m)l^2}\bigg(1+\frac{3b^2\beta^2}{r_+^2}(2-z_m)\bigg)
\bigg(\frac{T_c^3}{T^3}\bigg)\bigg(1-\frac{T^3}{T_c^3}\bigg).
\end{eqnarray}

According to the AdS/CFT dictionary, we have
$
\langle{\cal O}_+\rangle=lD_+r_+^{\Delta_+}l^{-2\Delta_+}.
$
Using the Eqs.~(\ref{D}) and (\ref{ct2:psi1}), the expectation value $<\mathcal{O}_{+}>$ is
\begin{equation}\label{critical exponents}
\frac{\langle{\cal O}_+\rangle^{\frac{1}{\Delta_+}}}{T_c}=\eta\frac{T}{T_c}
\bigg[\bigg(\frac{T_c^3}{T^3}\bigg)\bigg(1-\frac{T^3}{T_c^3}\bigg)\bigg]
^{\frac{1}{2\Delta_+}},
\end{equation}
with
\begin{eqnarray}\label{eta}
\eta&=&\frac{\pi}{z_m}\bigg\{\frac{2z_m}{2z_m+(1-z_m)\Delta_+}\left(
1+\frac{3\rho~m^2 l^2(1-z_m)}{8(3\rho-\kappa^2 \mu_0^3 l^2)}\right)\nonumber \\ &&\times \sqrt{\frac{1}{q^2(1-z_m)}
\bigg(1+\frac{3b^2\beta^2\mu_0}{\rho}(2-z_m)\bigg)}\bigg\}^{\frac{1}{\Delta_+}}.
\nonumber
\end{eqnarray}
From the Eq. (\ref{critical exponents}) we find that the
critical exponent is $1/2$,
which shows that the critical exponent is not affected by
the Gauss-Bonnet gravity, the Born-Infeld electrodynamics and the backreactions.

\section{Conclusion}

We have analytically investigated the properties of the holographic
superconductors for the Born-Infeld electrodynamics in the
Gauss-Bonnet gravity with backreactions near the phase transition.
We note that the analytical method is still available provided the
parameters $b$ and $\kappa$ are not too big, and the analytic
results are consistent with that obtained by the numerical
computation. The analytical expressions for the critical
temperature are obtained. From which we find that with the
increase of the Gauss-Bonnet parameter with fixed $b$ and
$\kappa$ the critical temperature decreases, which means
that the larger Gauss-Bonnet factor $\alpha$ makes the
condensation harder to form for the scalar field. As
the Born-Infeld parameter $b$ increases for fixed parameters
$\alpha$ and $\kappa$, the critical temperature decreases.
That is to say, the scalar condensation can be more difficult to form for larger
Born-Infeld parameter. With the increase of
the backreaction parameter $\kappa$ for fixed parameters
$\alpha$ and $b$, the critical temperature decreases,i.e.,
 the scalar condensation becomes difficult as the
backreaction parameter $\kappa$ increases. Furthermore, the Gauss-Bonnet parameter $\alpha$ modifies the
critical temperature more significantly than the backreaction parameter $\kappa$.
The effect of the backreaction parameter $\kappa$ on the critical temperature is bigger than the
Born-Infeld parameter $b$ by comparison. We also show
that the critical exponent is not affected by the Gauss-Bonnet
gravity, the Born-Infeld electrodynamics and the backreactions.

\begin{acknowledgments}

This work was partially supported by the National Natural Science
Foundation of China under Grant Nos. 11175065, 10935013;
the National Basic Research of China under Grant No. 2010CB833004;
PCSIRT, No. IRT0964; the Hunan Provincial Natural Science
Foundation of China under Grant No. 11JJ7001; and Construct
Program of the National Key Discipline.
\end{acknowledgments}

\end{document}